\documentclass{wscpaperproc}
\usepackage{latexsym}
\usepackage{graphicx}
\usepackage{mathptmx}
\usepackage{subcaption}
\usepackage{amsmath}
\usepackage{amsfonts}
\usepackage{amssymb}
\usepackage{amsbsy}
\usepackage{amsthm}
\usepackage{enumitem}

\usepackage[pdftex,colorlinks=true,urlcolor=blue,citecolor=black,anchorcolor=black,linkcolor=black]{hyperref}

\newtheoremstyle{wsc}
{3pt}
{3pt}
{}
{}
{\bf}
{}
{.5em}
{}

\theoremstyle{wsc}

\begin{document}

%
%

\pagestyle{fancyplain}

\thispagestyle{plain}
\firstPageHead{}

\chead{\fancyplain{}{\itshape Erazo, Goldsman, Keskinocak, and Sokol}}

\rhead{}
\cfoot{}
\renewcommand{\headrulewidth}{0pt} 

\makeatletter
\let\@internalcite\cite
\def\cite{\def\@citeseppen{-1000}%
    \def\@cite##1##2{(##1\if@tempswa , ##2\fi)}%
    \def\citeauthoryear##1##2##3{##1 ##3}\@internalcite}
\def\citeNP{\def\@citeseppen{-1000}%
    \def\@cite##1##2{##1\if@tempswa , ##2\fi}%
    \def\citeauthoryear##1##2##3{##1 ##3}\@internalcite}
\def\citeN{\def\@citeseppen{-1000}%
    \def\@cite##1##2{##1\if@tempswa, ##2)\else{}\fi}%
    \def\citeauthoryear##1##2##3{##1 (##3)}\@citedata}
\def\citeA{\def\@citeseppen{-1000}%
    \def\@cite##1##2{(##1\if@tempswa , ##2\fi)}%
    \def\citeauthoryear##1##2##3{##1}\@internalcite}
\def\citeANP{\def\@citeseppen{-1000}%
    \def\@cite##1##2{##1\if@tempswa , ##2\fi}%
    \def\citeauthoryear##1##2##3{##1}\@internalcite}
\def\shortcite{\def\@citeseppen{-1000}%
    \def\@cite##1##2{(##1\if@tempswa , ##2\fi)}%
    \def\citeauthoryear##1##2##3{##2 ##3}\@internalcite}
\def\shortciteNP{\def\@citeseppen{-1000}%
    \def\@cite##1##2{##1\if@tempswa , ##2\fi}%
    \def\citeauthoryear##1##2##3{##2 ##3}\@internalcite}
\def\shortciteN{\def\@citeseppen{-1000}%
    \def\@cite##1##2{##1\if@tempswa, ##2\else{}\fi}%
    \def\citeauthoryear##1##2##3{##2 (##3)}\@citedata}
\def\shortciteA{\def\@citeseppen{-1000}%
    \def\@cite##1##2{(##1\if@tempswa , ##2\fi)}%
    \def\citeauthoryear##1##2##3{##2}\@internalcite}
\def\shortciteANP{\def\@citeseppen{-1000}%
    \def\@cite##1##2{##1\if@tempswa , ##2\fi}%
    \def\citeauthoryear##1##2##3{##2}\@internalcite}
\def\citeyear{\def\@citeseppen{-1000}%
    \def\@cite##1##2{(##1\if@tempswa , ##2\fi)}%
    \def\citeauthoryear##1##2##3{##3}\@citedata}
\def\citeyearNP{\def\@citeseppen{-1000}%
    \def\@cite##1##2{##1\if@tempswa , ##2\fi}%
    \def\citeauthoryear##1##2##3{##3}\@citedata}
%
%
%
\def\@citedata{%
    \@ifnextchar [{\@tempswatrue\@citedatax}%
                  {\@tempswafalse\@citedatax[]}%
}

\def\@citedatax[#1]#2{%
\if@filesw\immediate\write\@auxout{\string\citation{#2}}\fi%
  \def\@citea{}\@cite{\@for\@citeb:=#2\do%
    {\@citea\def\@citea{, }\@ifundefined
       {b@\@citeb}{{\bf ?}%
       \@warning{Citation `\@citeb' on page \thepage \space undefined}}%
{\csname b@\@citeb\endcsname}}}{#1}}%

%
\def\@citex[#1]#2{%
\if@filesw\immediate\write\@auxout{\string\citation{#2}}\fi%
  \def\@citea{}\@cite{\@for\@citeb:=#2\do%
    {\@citea\def\@citea{; }\@ifundefined
       {b@\@citeb}{{\bf ?}%
       \@warning{Citation `\@citeb' on page \thepage \space undefined}}%
{\csname b@\@citeb\endcsname}}}{#1}}%

%
\def\@biblabel#1{}
\makeatother



\newdimen\bibindent
\bibindent=0.0em
\def\thebibliography#1{\section*{\refname}\list
   {}{\settowidth\labelwidth{[#1]}
   \leftmargin\parindent
   \itemindent -\parindent
   \listparindent \itemindent
   \itemsep 0pt
   \parsep 0pt}
   \def\newblock{}
   \sloppy
   \sfcode`\.=1000\relax}


\setlength{\baselineskip}{12.7pt}

\title{A SIMULATION-OPTIMIZATION FRAMEWORK TO IMPROVE THE ORGAN TRANSPLANTATION OFFERING SYSTEM}
\author{Ignacio Erazo\\
    David Goldsman\\
    Pinar Keskinocak\\
    Joel Sokol \\[12pt]
	H. Milton Stewart School of Industrial and Systems Engineering\\
	Georgia Institute of Technology\\
	755 Ferst Drive NW\\
	Atlanta, GA 30332-0205, USA\\}
\maketitle
\section*{ABSTRACT} {We propose a simulation-optimization-based methodology to improve the way that organ transplant offers are made to potential recipients.  Our policy can be applied to all types of organs, is implemented starting at the local level, is flexible with respect to simultaneous offers of an organ to multiple patients, and takes into account the quality of the organs under consideration.  We describe in detail our simulation-optimization procedure and how it uses data from the Organ Procurement and Transplantation Network and the Scientific Registry of Transplant Recipients to inform the decision-making process. In particular, the optimal batch size of offers is determined as a function of location and certain organ attributes. We present results using our liver and kidney models, where we show that, under our policy recommendations, more organs are utilized and the required times to allocate the organs are reduced over the one-at-a-time offer policy currently in place}.

\section{INTRODUCTION}
\label{sec:intro}

Organ transplants have allowed many potential transplant patients to extend and improve the quality of their lives by giving them the chance of receiving a functional organ. In 1984, the Organ Procurement and Transplantation Network (OPTN) was created with the goal of overseeing the transplantation system in the United States. As of March 2022, there were more than 105,000 candidates on the national waiting list, and every ten minutes a new candidate is added to the list \shortcite{[1]}. Even though the numbers of donors and transplants have increased steadily over the years, there is still a significant shortage of organs; about 20 people die each day while waiting for an organ \shortcite{[2]}. The United States had 36.88 deceased organ donors per million population (PMP) (the second-highest rate in the world) and 22.99 living donors PMP (7th) in 2019 \shortcite{[3]}. Unfortunately, a significant percentage of organs received by the OPTN for transplant are discarded, contributing to the demand-supply gap. For example, 19\% of kidneys are discarded, while 9\% of livers are discarded \shortcite{[4]}. The United Network for Organ Sharing (UNOS)---the institution that administers the OPTN---has recently implemented several changes in its decision-making process with the goal of improving organ utilization \shortcite{[5]} (i.e., the proportion of organs ultimately transplanted out of the total number of organs suitable for transplantation).

Currently, after procuring an organ, OPTN generates a prioritized list of patients according to their models of compatibility and the patients’ health markers. One-at-a-time sequential offers are made to the respective transplant teams representing the patients, according to the patients'  placements in the priority list for the organ, until the organ is accepted or until it becomes inviable. In the past, teams did not have a time limit to provide an answer to the offer; but nowadays offers expire after one hour for the first (primary) transplantation center and after 30 minutes for all the subsequent transplantation centers to which an offer is made \shortcite{[6]}. The creation of the prioritized patient list depends on several factors and varies depending on the organ being offered. In fact, several articles have studied how to create “optimal” prioritized offering lists \shortcite{[7],[9],[8]}. 

Over the past few years, simulation has been widely used to study a variety of healthcare problems. For instance, optimizing the allocation of resources \shortcite{Kilinc2020} and guiding policy-making \shortcite{Moustaid2019}. In the context of the transplantation system, simulation has been applied to propose and evaluate new models \shortcite{Sandi2019},  study the roles of incentives \shortcite{Konrad2020}, and  test new policies \shortcite{[10]}.

Our goal is to improve the organ offering system.  That is, 
given a particular prioritized list of organ transplant candidates, we will propose and study improvements to the process by which organs are actually offered to those patients. It is widely known that increasing the number of donated organs that are actually transplanted can result in great benefits to society; but there are also certain costs involved in the offering process that need to be taken into consideration. That being said, we propose a simulation-optimization-based methodology that helps to create a policy that maximizes the overall “gain” (defined as benefits minus costs) accrued by the organ-allocation system during the offering process. We will use simulation to evaluate the ``gain'' performance of several different policies using data from the SRTR and the OPTN\@. Careful examination of the simulations' results allows us to determine a policy that assigns the number of offers to make simultaneously in batches for each type of organ (i.e., kidney, liver, etc.)\ that takes into consideration the characteristics of the organ and the location where the organ arrived. Our contributions are three-fold: 
\begin{enumerate}[label=(\roman*)]
    \item We propose a simple, flexible, yet accurate simulation model that works for all organs and that starts at the local level, unlike \shortcite{Sandi2019} who assess the system at a national level. 
    \item We consider batch-offering policies as do \shortcite{[10]}; however, our batch sizes are not fixed in advance, and our policies also incorporate organ characteristics and the location of the organs into the decision-making process.
    \item We compare the quality of organs donated under our new policy to the overall quality of organs donated in the United States, which helps to ease concerns with respect to the use of organs that are otherwise lost.
\end{enumerate}

The rest of the article is organized as follows: Section \ref{sec:Model} gives the model overview; specifically, Section \ref{subsec:Data} explains data considerations, and Section \ref{subsec:Procedure} details our simulation-optimization procedure. Afterwards, Section \ref{sec:ModelVal} presents our model's validation (compared to the SRTR and OPTN data), and Section \ref{sec:Experiments} describes our experimentation and the performance of our proposed policy. Finally, Section \ref{sec:Discussion} discusses the results, and Section \ref{sec:Conclusions} provides conclusions.

\section{MODEL OVERVIEW}\label{sec:Model}

The offering policy encountered in actual practice is very complex; however, in most cases, offers are first made for patients having the highest priority on the waiting list within the OPO/DSA (organ procurement organization/donation service area) where the organ arrives. If no patient within that area accepts an offer to take that organ, then the organ is offered to the highest-ranked patient on the regional priority list (each region is formed by one or more OPOs, and each OPO is constituted by one or more transplantation centers); and if none accept in that region, then the organ is offered nationally. For simplicity, in the rest of this work we will consider this offering structure, just as \shortciteN{[10]} did. Our goal is to maximize the overall “gain” of the transplantation system, where we consider a ``profit/cost'' structure for which a successful organ transplant results in a profit; making the  offers themselves incurs a small cost; and, moreover, there is a significant disappointment cost for a patient who accepted an offered organ but did not actually end up receiving that organ (this can happen since we are making multiple offers simultaneously in order to increase the chance of a successful allocation). 

\subsection{Data}\label{subsec:Data}

This study used data from the Scientific Registry of Transplant Recipients (SRTR). The SRTR data system includes data on all donor, wait-listed candidates, and transplant recipients in the US, submitted by the members of the Organ Procurement and Transplantation Network (OPTN)\@. The Health Resources and Services Administration (HRSA), U.S. Department of Health and Human Services provides oversight to the activities of the OPTN and SRTR contractors.

In fact, several data sources were used in order to compute all of the needed parameters for the simulation model. Public data from the 2018--2019 time period involving the waiting list (removals, new candidates, and average length), transplants performed, and expected acceptance ratio of organs per transplantation center was obtained from the SRTR \shortcite{[4]}. Private OPTN data from the time period 2012--2014 involving the pool of organs donated and organs lost was also used to evaluate the quality of organs (that time period was used because it is the last to include that information among the files shared under the collaboration agreement with the OPTN)\@. Finally, the maximum cold-ischemia time (CIT, i.e., the time between the chilling of a tissue, organ, or body part after its blood supply has been reduced or cut off and the time it is warmed by having its blood supply restored) considered was 16 hours for livers and 40 hours for kidneys; these values are consistent with the highest CITs encountered in practice, according to \shortciteN{[11]} and \shortciteN{[12]}. We now discuss preparation of organ data, followed by that of transplantation center data.

\subsubsection{Preparation of Organ Data}
We first detail the process of estimating the probability distribution of the arrival locations for the organs. Since most of the organs are transplanted in the same OPO/DSA where they are initially offered, a reasonable approximation of the probability of a certain organ arriving to a specific OPO/DSA given that it entered the transplantation system can be computed by the number of transplants successfully undertaken in that DSA, divided by the total number of successful transplants in the country.

We considered various categories representing different characteristics of the organs that were used, classified separately by liver and kidney. These categories were selected according to what is presented in the SRTR reports \cite{[4]}, and are: ``Donor after cardiac death'' (DCD), ``Hepatitis C virus positive'' (HCV+), and “Normal” for livers; and “KDRI\,$<$\,1.05”, “1.05\,$\leq$\,KDRI\,$<$1.75”, and “1.75\,$\leq$\,KDRI” for kidneys, where the Kidney Donor Risk Index (KDRI) estimates the relative risk of post-transplant kidney graft failures depending on the characteristics of the deceased donor versus a reference deceased donor. For each organ and respective category, the total number of transplants carried out was divided by the utilization rate (percentage of organs that were actually transplanted) to obtain the total arrivals. Afterwards, the expected number of organs arriving at each center per year was computed by averaging the number of transplants per center (during 2018--2019), normalizing it (with respect to the United States), and then using that proportion multiplied by the total number of organs retrieved in the country (per category) as the estimate. For each type (kidney, liver) and category of organ, we used the data involving accepted organs per OPO and the total number of organs of each type and category that arrived to the country's pool to obtain the location probability distributions, i.e., the probability distributions of the locations (OPOs) of an arriving organ of a specific type and category.

All the discarded organs during the 2012--2014 period were analyzed, and after excluding all organs that were not considered apt for transplant (e.g., an organ deemed too old, poor biopsy findings, structural damage, potential sicknesses from the donor, among other reasons) we estimated how many organs were lost because of not being accepted before the maximum CIT was exceeded. Among discarded kidneys, 19.9\% were lost because the maximum CIT was exceeded, while only 9.4\% for the livers. The quality of those organs was compared to the quality of the organs transplanted in that period by looking at relevant health markers of the donors.

\subsubsection{Preparation of Center Data}

Waiting list information was used to estimate the average number of removals, new candidates, and average length of the waiting list. The last value was used to create starting conditions for the simulation, while the other two were used to update the waiting list during the simulation.

In addition, for each category of organ (i.e., “DCD” liver, “HCV+” liver,\ldots) we needed to compute the probability of acceptance at a given center. For each category, using the nationwide data on offers, the probability of an offer being accepted in the United States was computed by dividing the total number of accepted offers for that category by the total number of offers made. Then, for each center and category, we multiplied that probability by the offer acceptance ratio (with respect to the U.S. average) that the (center, category) pair had according to \citeN{[4]}. The use of that statistic is relevant because it considers the attributes of the organs that were offered as well as the characteristics of the potential recipients to whom the organs were offered.  The computation of that offer acceptance ratio is carried out and reported by the SRTR.

\subsection{Procedure}\label{subsec:Procedure}

Using the data described previously, we built a simulation model that can simulate one year of the transplantation system for any organ (by using the correct input parameters) under different policies. The simulation model returns as output the total ``gain'' of the system (defined below) over the one-year period.

In order to obtain an optimal policy, we take the following simulation-optimization approach: For different ``batch sizes'' $x$, we perform multiple simulation replications of the behavior of the transplantation system for each organ separately under the policy of making batches of ``$x$'' simultaneous offers for \emph{all organs arriving to the system}. During each simulation we keep track of the ``gain'' of all the organs that enter the system, where the term ``gain'' denotes the result of a profit/expense point system that awards points for successful transplants, but deducts points based on transplant team decision effort as well as disappointed potential recipients who ask for an organ that is ultimately awarded to someone else.  In any case, at the end of this phase, we obtain a ``training set'' (derived from all the simulations) where the information vectors represent the organs that have arrived into the system via the triples (Category, OPO, policy ``$x$''), and the response vector represents the ``gain'' of any specific organ when it was offered in the system. We optimize over the policies for each pair (Category, OPO) in the ``training set'' by finding the value $x^\star$ among the possible policies ``$x$'' that maximizes the average gain for that particular pair (Category, OPO) over the ``training set''; and as such, we obtain our proposed policy that aims to optimize the ``gain'' of the transplantation system. It is clear that our simulation-optimization procedure allows for different numbers of simultaneous offers for the different (Category, OPO) pairs, which is novel when compared to the current literature, where policies specify a fixed number of offers to be given for each organ, regardless of the (Category, OPO) pair.

\subsubsection{Simulation Initialization}

To simulate one year of the transplantation system for each type of organ the following steps are executed:
\begin{itemize}
    \item We create each center and we initialize them with a waiting list of potential organ recipients.  The length of this waiting list is within 20\% of the average size of the waiting list that the center has experienced over time, and is computed by using a uniform integer random variable in the interval between 80\% and 120\% of the average waiting list size computed in the data preparation step. We chose the 80\% and 120\% bounds so as to mimic the variability  exhibited by the SRTR/OPTN from year to year over different centers.
    \item We determine the number of new listings for each center during the year and the number of removals (non-transplants) associated with each center during the year. These values are also generated by taking uniform random integers in the interval [80\%,120\%] of the averages obtained in the data preparation step---again, the 80\%, 120\% bounds are motivated by the historical data. These system arrivals and departures are considered to be distributed uniformly over the 365 days of the year. 
    \item We determine the number of organs that arrive at the system in one year for each category of the respective organ type. Once more, we do so by selecting uniform random integers between 0.8 and 1.2 times the average numbers of organs that arrived in each category nationwide; this design choice provides variability and robustness with regard to the total organs donated, which may be necessitated by changes in donation patterns, as well as external events (such as COVID)\@. Then, each organ created in the simulation is mapped to an OPO depending on the probability distribution of the location of that specific category of organ. The time of the organ arrival is also uniform over the year where this choice was made due to observed consistency of monthly donation data, but also a lack of sufficient data to create separate distributions over days/weeks.
    \item Finally, each organ (depending on its category) has a certain probability of being discarded for medical reasons. This way we ensure that all the organs offered are from the population of legitimate organs that can be transplanted. 
\end{itemize}

\subsubsection{Simulation Logic}

\begin{figure}[h!]
    \centering
    \includegraphics[scale=.65]{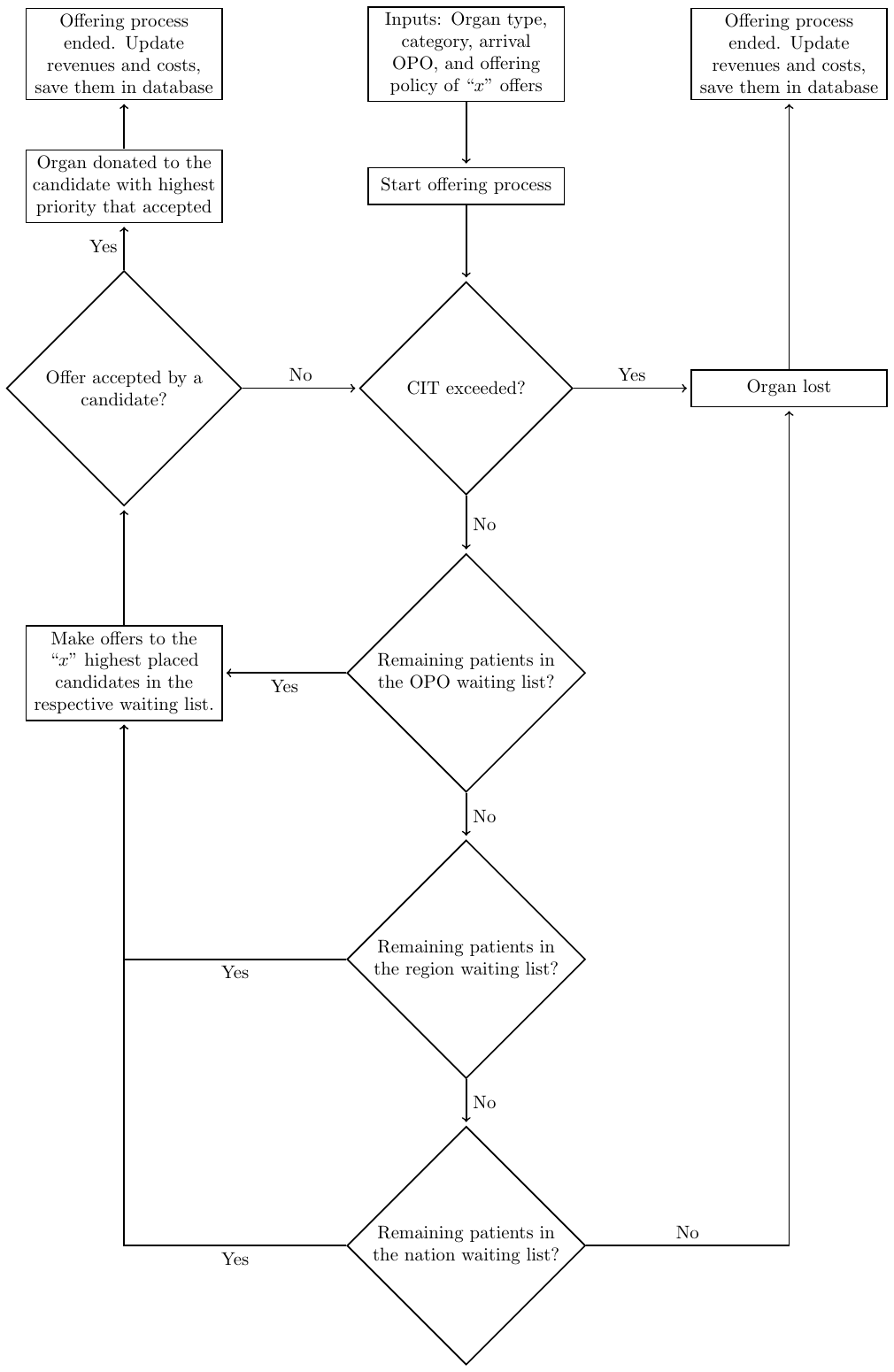}
    \caption{Offering procedure.}
    \label{fig:OfferingProcedure}
\end{figure}

\begin{figure}[h!]
    \centering
    \includegraphics[scale=.65]{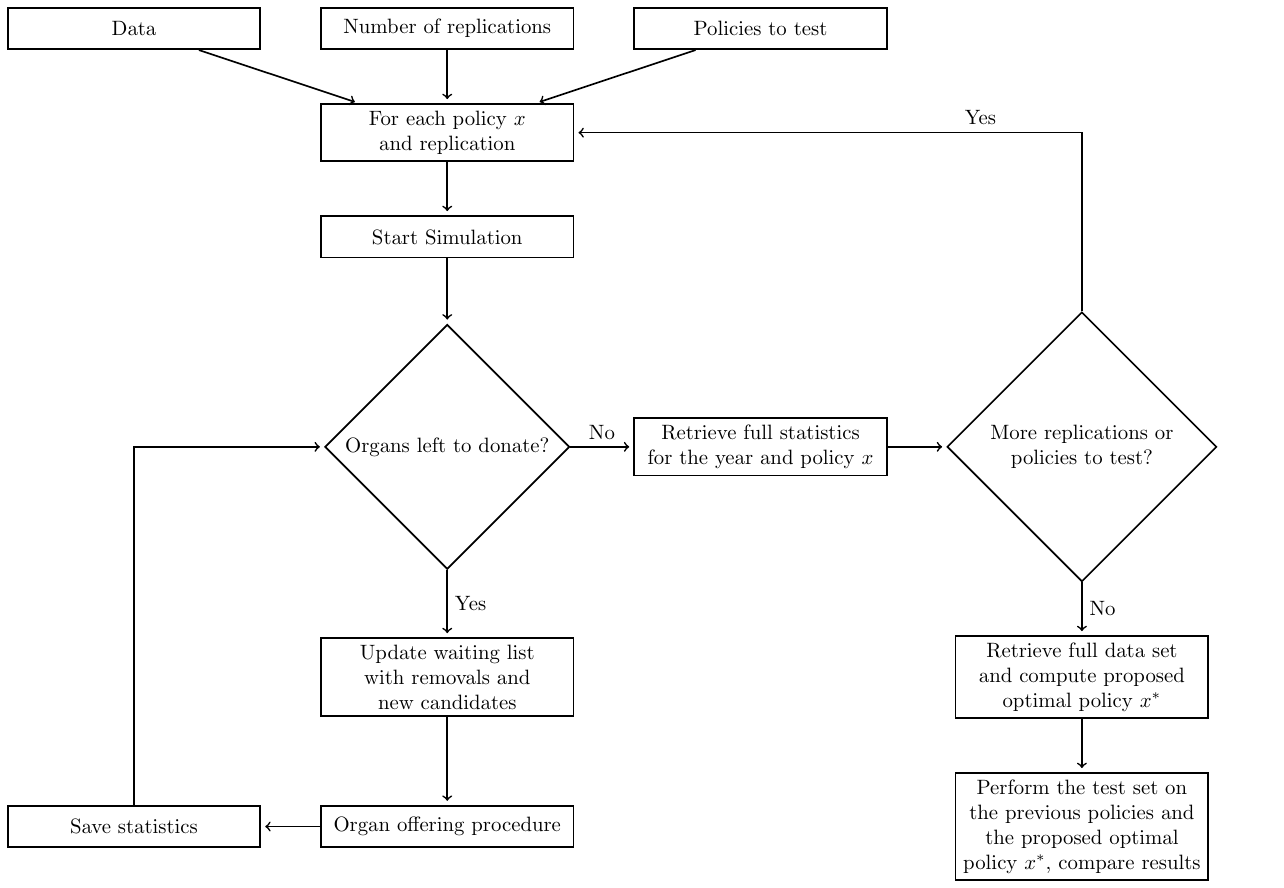}
    \caption{Simulation flowchart.}
    \label{fig:SimulationProcedure}
\end{figure}

All the events (new patients arriving to the centers' waiting lists, removals, and organs arriving) are added to an event queue, and the simulation is executed by jumping discretely from event to event in time order. In particular, every time an organ arrives, the offering procedure presented in Figure \ref{fig:OfferingProcedure} is executed. Figure \ref{fig:SimulationProcedure} is a simple flowchart that depicts most of the simulation process.

{}

{}

Due to a lack of data, we specified that the first time a center receives an offer for a specific organ, it will take Unif(10,60) minutes to give an answer (i.e., to decide if the patient to whom the organ was offered accepts or not). The upper bound was a known value that follows from the actual system, whereas the distribution was selected to be uniform and the lower bound was set  via expert opinion, and then tested experimentally and tuned so as to make the model consistent (see the validation discussion in Section \ref{sec:ModelVal}). After that, as the center's team already knows the specifics of the organ, the time needed by the team to give subsequent responses for that organ (should it not be accepted in the first round) will likely be smaller.  For such subsequent responses, we assumed a uniform response time of between 0 (an “automatic” answer) and 10 minutes for both livers and kidney, where the lower bound is known (because of automatic rejection) and the upper bound was tested/tuned experimentally and based on expert opinion. Again, the choice of these parameters is backed up by the model validation discussed in the next section.

\section{MODEL VALIDATION}\label{sec:ModelVal}

We tested the behavior of our simulation model against the SRTR data (from 2018) with the goal of validating that it accurately reflects what happens in practice. We conducted 50 replications of one year for livers and kidneys, and the comparisons versus the real data are presented in Tables \ref{ResultsValidationLiver} and \ref{ResultsValidationKidney}, respectively (where the entries for allocated organs, total offers, and minutes to allocate are rounded to integers). Note that our ``batching'' parameter was set to $x=1$, corresponding to the actual 1-offer-at-a-time policy. 
	\begin{table}[h!]
	\footnotesize
		\caption{Validation of the liver model versus SRTR Data.}
			\label{ResultsValidationLiver}
			\centering
		\begin{tabular}{||c||c|c|}
			\hline
			& 1-Offer Simulation & SRTR Data 2018 \\
			\hline
			Organ Utilization & 90.95\% & 91.10\%\\
			\hline
			Allocated Organs & 6,872 & 7,003\\
			\hline
			Total Offers & 168,109 & 168,159 \\
			\hline
			Minutes to Allocate & 202 & --- \\
			\hline
		\end{tabular}
			\end{table}

	\begin{table}[h!]
	\footnotesize
		\caption{Validation of the kidney model versus SRTR Data.} 
	\label{ResultsValidationKidney}
				\centering
		\begin{tabular}{||c||c|c|}
			\hline
			& 1-Offer Simulation & SRTR Data 2018 \\
			\hline
			Organ Utilization & 80.80\% & 81.00\%\\
			\hline
			Allocated Organs	& 13,829 & 13,752\\
			\hline
			Total Offers & 1,563,116 & 1,562,014 \\
			\hline
			Minutes to Allocate	& 605 & --- \\
			\hline
		\end{tabular}
		\end{table}

All of the simulated values are within 2\% of the real values reported by the SRTR in 2018, which helps to validate the choice of parameters and also to establish that our model accurately represents the actual behavior of the offering system for the respective organs.

\section{EXPERIMENTS}\label{sec:Experiments}

For both liver and kidney, we performed our simulation-optimization procedure with the goal of obtaining a policy maximizing the ``gain'' of the transplantation system (for that particular organ). For the liver model, the ``training set'' was created with 50 simulations of one year of operations of the transplantation system for each policy of “$x$” simultaneous offers, where $x$ was considered for values of $1, 2, \ldots, 30$. With respect to the kidney model, the ``training set'' was comprised of 10 one-year replications for each policy choice of $x =  1, 2, \ldots, 30$ simultaneous offers. The difference in the lengths of the training sets is due to the running times of each of the models --- the kidney model is slower because of having a much larger number of donated organs, more people on the waiting list, and a significantly greater probability of organs not being accepted, thus generating more iterations.

With respect to the cost parameters used, they are as follows: The system ``earns'' 1000 units when an organ is transplanted; the system incurs a cost of 25 units the first time a specific organ is offered to a transplantation center (personnel must expend time to review the organ as a team, etc.), then each subsequent offer for that same organ received by the same center (offered to another patient) only costs 1 unit (because the team already has previous data on the organ and needs less time/resources to decide about the next offer); and finally the disappointment cost of a patient not receiving an organ whose offer he/she has accepted is 300. These seemingly arbitrary earnings/costs reflect the extremely high value a successful transplant has, but also the negative consequences that a disappointed offer may have on the patients.

After optimizing over the ``training set'' to obtain an optimal policy $x^*$(Category, OPO) for each pair (Category, OPO), we tested the quality of our proposed policy $x^*$ by running new replications of the transplantation system (i.e., the ``testing set'', with 50 replications for liver, 10 for kidney), and we compared the results to those obtained by static policies having $x = 1, 2, \ldots,15$ simultaneous offers for all organs regardless of the respective (Category, OPO) pair.

\subsection{Results for the Liver Model}

\begin{figure}[h]
\begin{center}
\begin{subfigure}{.48\textwidth}
  \centering
  \includegraphics[width=0.98\linewidth]{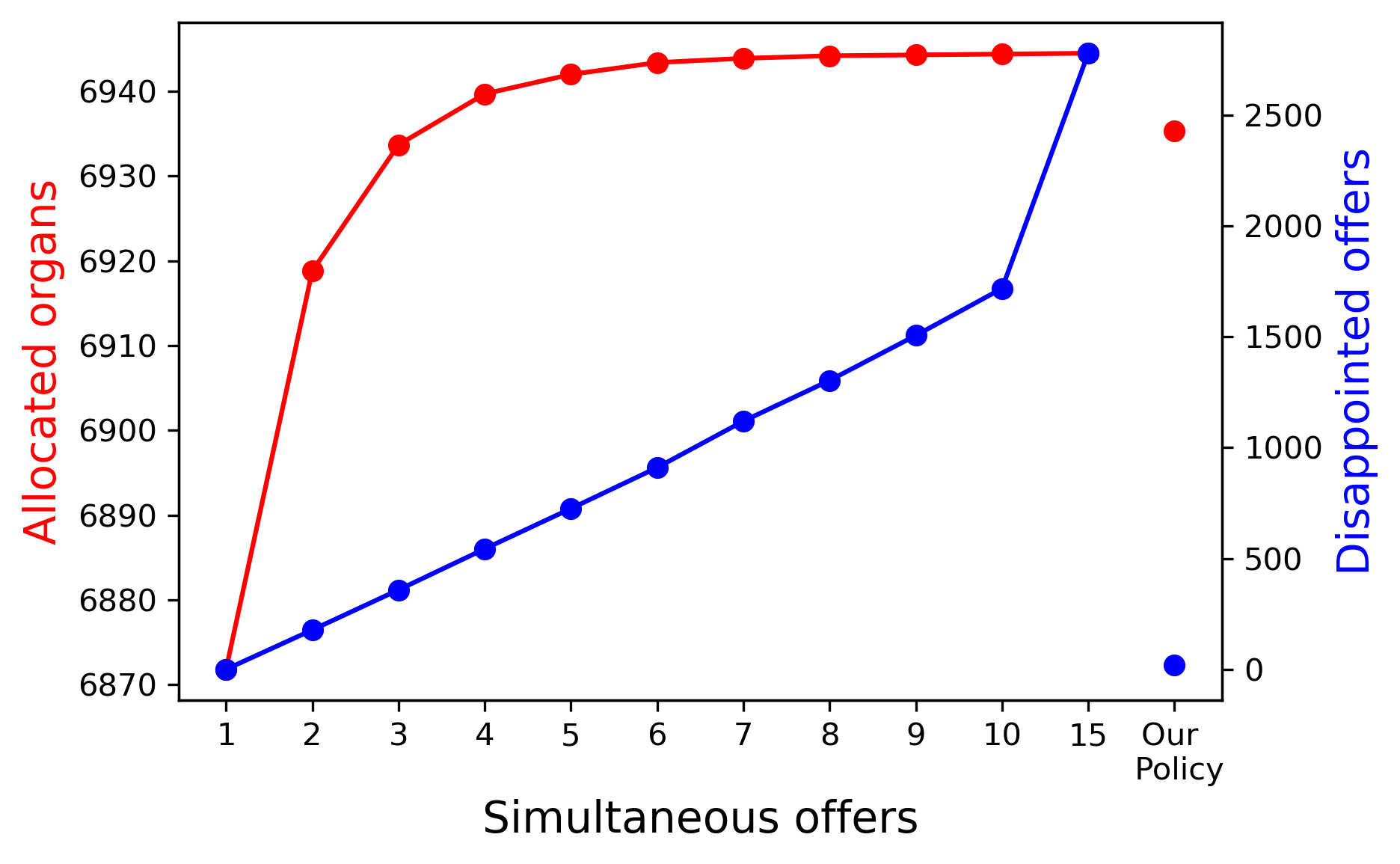}  
  \caption{Number of allocated organs and disappointed offers per policy for the liver model.}
  \label{fig:LiverDonations}
\end{subfigure}
\begin{subfigure}{.45\textwidth}
  \centering
  \includegraphics[width=0.98\linewidth]{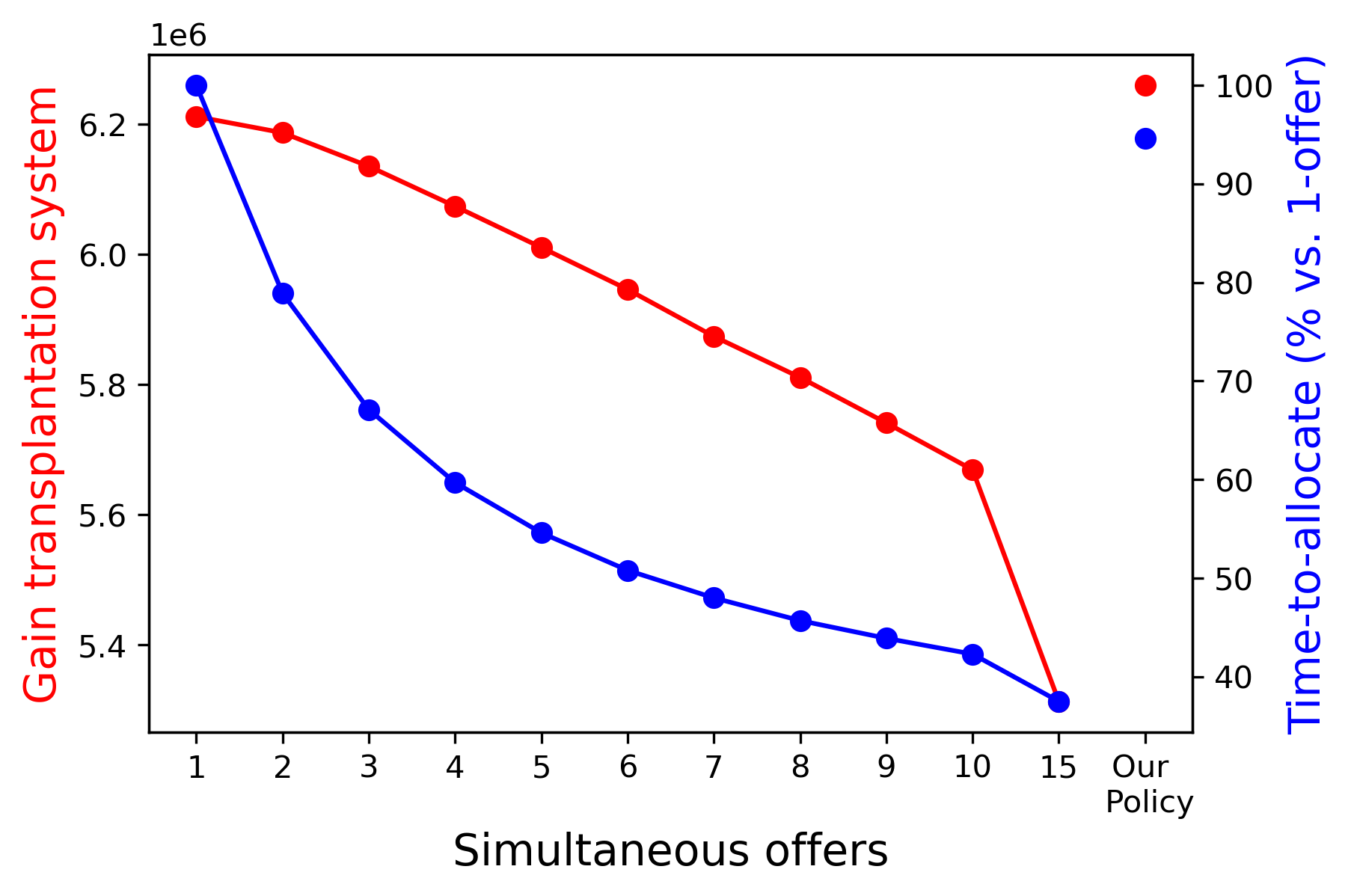}  
  \caption{Gain of the transplantation system and time-to-allocation per policy for the liver model.}
  \label{fig:LiverRevenues}
\end{subfigure}
\caption{Results of the test set for the liver model.}
\end{center}
\end{figure}

Figure \ref{fig:LiverDonations} depicts the total number of allocated organs and the total number of disappointed offers for a subset of the policies ($x = 1,2,\ldots,10,15$ and our proposed policy); and Figure \ref{fig:LiverRevenues} graphs the total ``gain'' of the transplantation system and the time-to-allocation (versus the one-offer policy) for that subset of policies. The one-offer-at-a-time policy has a utilization rate of 90.95\%, whereas the policy with 15 simultaneous offers has an organ utilization of 91.92\%. Our proposed policy, which uses a different number of offers for each organ depending on its (OPO, Category) pair, with that number coming from the optimization over the training set results, has an organ utilization of 91.79\%.

Our proposed policy outperforms the current policy with respect to the average number of allocated organs by about 63 additional organs per year, and also reduces by 5.3\% the time needed to allocate the organs. The proposed strategy also outperforms the policies with 2 simultaneous offers and with 3 simultaneous offers with respect to allocated organs and revenues, while making fewer overall offers. Compared to the proposed policy, the policy of 4-at-a-time offers is the first yielding more allocated organs, with the bonus of reducing the time to allocate by 37\%; however, those advantages come at the cost of increasing by a factor of 25 (543.5 vs.\ 20.5) the number of disappointed offers (where a patient accepted the offer but the transplant was not assigned to that patient). Similarly, the strategy of using batches of 15 offers increases the number of organs transplanted by just 73 vs.\ the current benchmark one-at-a-time batch policy, while achieving allocations 60\% faster, but with an increase of more than 2,600 offers disappointed, and more than 50,000 extra offers. It is clear that when the number of simultaneous offers increases, the gain of the transplantation system decreases almost linearly. In any case, we see that the highest overall gain for the system under the current ``gain'' and cost parameters is given by the proposed policy. 

We briefly discuss the quality of the ``extra” livers that end up being transplanted above and beyond what the current benchmark policy would have achieved.  All of these organs come from the pool of livers that would have been discarded under the current policy due to exceeding the CIT constraint. Table \ref{ComparisonDataOrgansLiver} compares several statistics between this pool and the livers that were actually donated under the current policy. The two sets of livers (donated and discarded due to CIT) have similarities with respect to the age of the deceased donors and the percentage of donors having diabetes or hypertension. On the other hand, there is a significant difference regarding the proportion of DCD donors. 

	\begin{table}[h!]
	\footnotesize
	\caption{Health markers for donated livers and livers discarded due to CIT.}
		\label{ComparisonDataOrgansLiver}
		\centering
		\begin{tabular}{||c||c|c|}
			\hline
			& CIT Discarded Livers & Donated Livers \\
			\hline
			DCD donors (\%) & 78.8\% & 4.5\%\\
			\hline
			Donors with diabetes (\%)	& 9.6\% & 10.7\% \\
			\hline
			Donors with hypertension (\%) & 74.9\% & 66.8\% \\
			\hline
			Average Body Mass Index (BMI) of donors	& 26.8 & 26.8 \\
			\hline
			Average Age of donors (years)	& 37.8 & 39.5 \\
			\hline
			Age $<$ 18 (\%)	& 8.7\% & 9.6\% \\
			\hline
			18 $\leq$ Age $\leq$ 34 (\%)	& 37.5\% & 31.9\% \\
			\hline
			35 $\leq$ Age $\leq$ 49 (\%)	& 28.8\% & 25.5\% \\
			\hline
			50 $\leq$ Age $\leq$ 64 (\%)	& 20.7\% & 25.5\% \\
			\hline
			65 $\leq$ Age (\%)	& 4.3\% & 7.5\% \\
			\hline
		\end{tabular}
		
	\end{table}


\subsection{Results for the Kidney Model}

\begin{figure}
\begin{center}
\begin{subfigure}{.48\textwidth}
  \centering
  \includegraphics[width=0.98\linewidth]{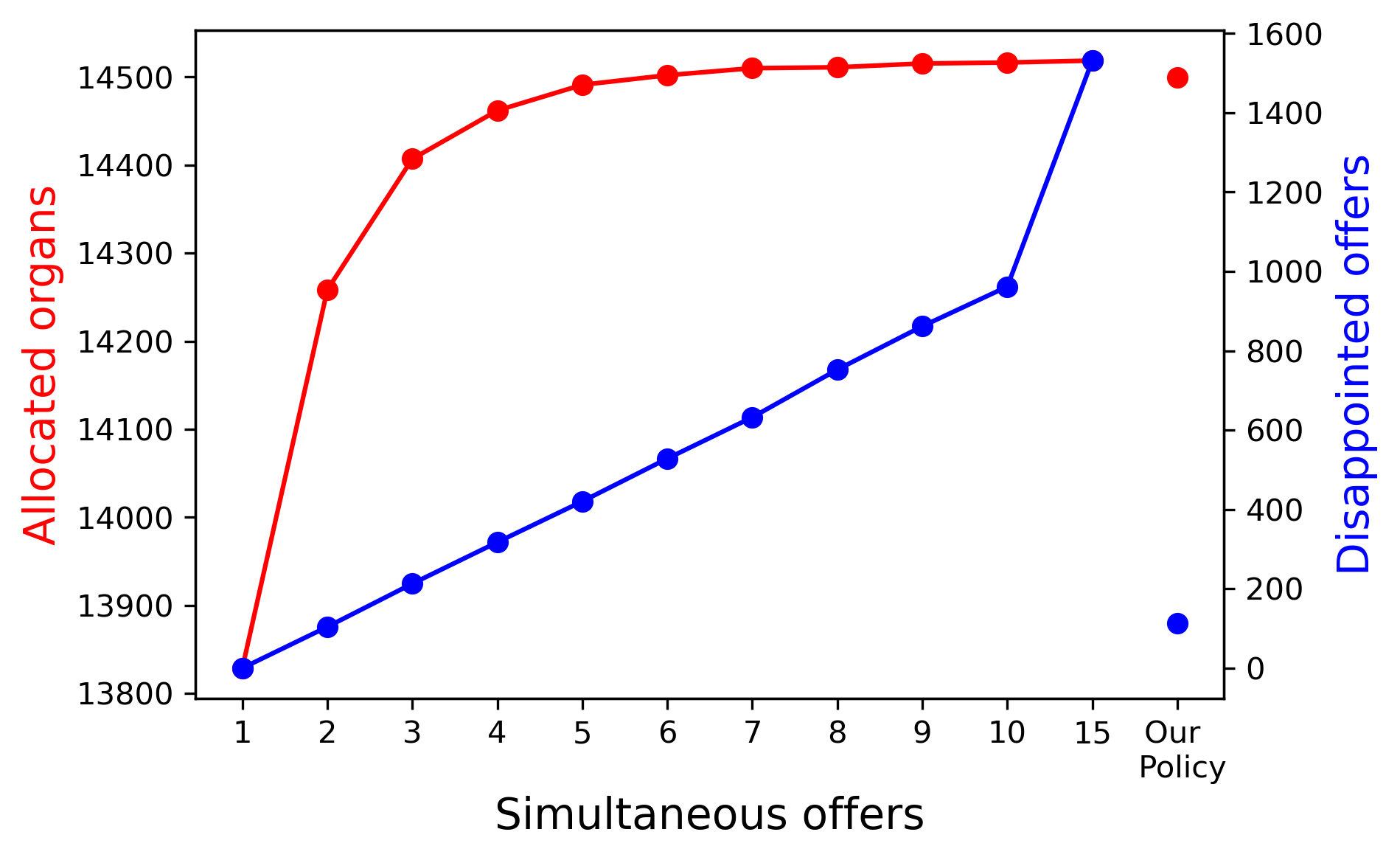}  
  \caption{Number of allocated organs and disappointed offers per policy for the kidney model.}
  \label{fig:KidneyDonations}
\end{subfigure}
\begin{subfigure}{.45\textwidth}
  \centering
  \includegraphics[width=0.98\linewidth]{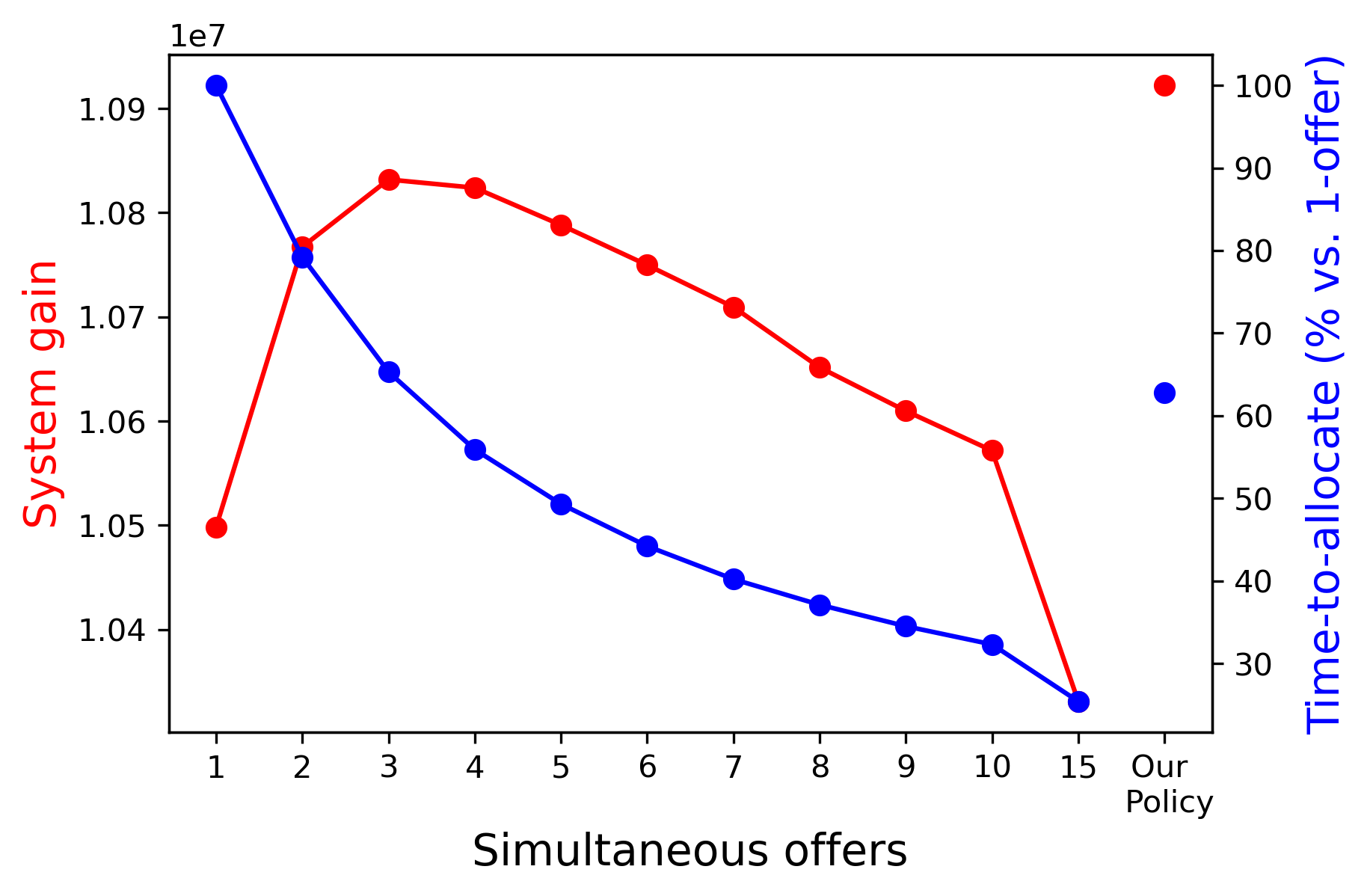}  
  \caption{Gain of the transplantation system and time-to-allocation per policy for the kidney model.}
  \label{fig:KidneyRevenues}
\end{subfigure}
\caption{Results of the test set for the kidney model.}
\end{center}
\end{figure}

Figure \ref{fig:KidneyDonations} illustrates the total number of allocated organs and the total number of disappointed offers for a subset of the policies ($x = 1,2,\ldots,10,15$ and our proposed policy); and Figure \ref{fig:KidneyRevenues} shows the total ``gain'' of the transplantation system and the time-to-allocation (versus the one-offer policy) for that subset of policies. The benchmark one-offer-at-a-time scenario has an average organ utilization of 80.80\%, whereas the policy with 15 simultaneous offers has an organ utilization of 84.80\%, and our proposed policy has an organ utilization of 84.68\%. The latter policy outperforms the benchmark policy with respect to the average number of allocated organs, as there is an increase exceeding 650 organs. Furthermore, the time needed to allocate the organs is reduced by 37.2\%. The proposed strategy also outperforms all of the policies with at most 5 simultaneous offers with respect to allocated organs and revenues. Besides that, the proposed policy  results in fewer disappointed offers than most of the other policies. For the sake of comparison, the strategy with batches of size 15 offers increases the number of transplanted kidneys by around 20 per year, and reduces the amount of time to allocate those organs by 74.6\% (compared to the one-offer policy); but it causes 1,417 extra disappointed offers compared to our proposed policy. The 6-offers policy allocates only 2 extra organs (compared to our proposed strategy) but does so with 415 extra disappointed offers and with 18,000 more total offers.  

Table \ref{ComparisonDataOrgansKidney} provides a comparison between “extra” kidneys that end up being transplanted above and beyond what the current benchmark one-at-a-time batch policy would have achieved. All of these organs come from the pool of kidneys that would have been discarded under the current policy due to exceeding the CIT bound. Kidneys discarded due to exceeding the CIT bound tend to come from older deceased donors, have a greater frequency of being from DCD donors, and have a higher rate of diabetic donors (correlated with the higher age); but, those discarded kidneys have a lower hypertension frequency.
	\begin{table}[h!]
	\footnotesize
		\caption{Health markers for donated kidneys and kidneys discarded due to CIT.}
		\label{ComparisonDataOrgansKidney}
		\centering
		\begin{tabular}{||c||c|c|}
			\hline
			& CIT Discarded Kidneys & Donated kidneys \\
			\hline
			DCD donors (\%) & 20.5\% & 13.9\%\\
			\hline
			Donors with diabetes (\%)	& 24.5\% & 7.2\% \\
			\hline
			Donors with hypertension (\%) & 38.4\% & 73.3\% \\
			\hline
			Average Body Mass Index (BMI) of donors	& 28.3 & 27.0 \\
			\hline
			Average Age of donors (years)	& 53.1 & 37.3 \\
			\hline
			Age $<$ 18  (\%)	& 1.4\% & 11.2\% \\
			\hline
			18 $\leq$ Age $\leq$ 34 (\%)	& 9.8\% & 33.2\% \\
			\hline
			35 $\leq$ Age $\leq$ 49 (\%)	& 22.0\% & 27.9\% \\
			\hline
			50 $\leq$ Age $\leq$ 64 (\%)	& 46.8\% & 24.4\% \\
			\hline
			65 $\leq$ Age (\%)	& 20.0\% & 3.3\% \\
			\hline
		\end{tabular}
	
	\end{table}

\section{DISCUSSION}\label{sec:Discussion}

With respect to livers, our results suggest that by increasing the number of simultaneous offers, more transplants would take place, and thus more lives could be saved; and, in addition, outcomes should improve by reducing the CITs of the transplanted organs \shortcite{[11]}. It is clear that a new policy that makes it easier to use these otherwise discarded organs will provide the recipients with organs of similar quality except the fact that they often come from DCD donors. In fact, several countries have apparently had good results with organs from DCD donors, as evidenced by a study evaluating post-transplantation outcomes in more than 1,000 DCD and non-DCD livers donated, with an insignificant difference in patient survival \shortcite{[13]}, and by the evidence of non-inferiority of DCD transplants found in \shortciteN{[14]}. Thus, the donation of such discarded organs may be feasible and useful for society. 

Now considering kidneys, even though the overall quality of discarded kidneys seems to be lower, countries such as Spain routinely and successfully use organs donated from patients over 70 years old, where 23\% of donations come from the 70–79 age group, and 10\% come from the over-80 age group, according to recent data \shortcite{[15]}. This suggests that a significant number of the discarded organs are of sufficient quality to provide good results for the recipients. Overall, our proposed policy achieves the highest net gain for the transplantation system, and does so by increasing the number of donated organs while barely increasing the number of disappointed offers.

\section{CONCLUSIONS}\label{sec:Conclusions}
Our model was designed to replicate the behavior of the organ transplantation system in the United States, and the goal was to maximize the overall “earnings'' obtained by society. Our simulation-optimization procedure allowed us to obtain a policy that would achieve that goal by specifying the number of offers to make for each individual organ that arrives to the system, depending on its category (e.g., “normal”, DCD, or HCV+ for livers) and its location when it arrives to the system. The model proved to be robust with respect to the organ it was used on, and is actually applicable to any organ as long as the appropriate parameters (i.e., arrival rates, acceptance probabilities, etc.) are used as inputs. 

Because of the “earning”-maximization objective undertaken in our framework, our policy is guaranteed to have the near-best earnings for the system; and for different cost parameters where the ``revenues'' for donating an organ exceed the costs of disappointed offers, our proposed strategy yielded better organ utilization than the current system as well as faster times-to-allocate; and, moreover, the number of extra organs transplanted greatly exceeded the number of disappointed offers. Compared to other policies using fixed, small batch sizes, our ``optimal'' policy also yielded increased earnings, and either resulted in a larger number of organs transplanted, or had a bit fewer but with just a fraction of the disappointed offers. This shows that our policy provides a good balance between organs transplanted and disappointed offers. Our results also suggest that the organ utilization of the system can be greatly increased by making multiple offers; and even though the overall quality of those otherwise discarded organs is lower, empirical evidence from previous studies suggests that the outcomes from such organs are nearly identical and clearly good enough to provide benefits for the recipients and save lives. Overall, we would expect that more lives could be saved because of the ability to transplant more organs. In addition, as the time-to-transplant is reduced, better outcomes for the recipients can be expected \shortcite{[16],[17]}.

With respect to the cost parameters we used, note that the cost of disappointed people is high (e.g., the earnings obtained by donating one organ will be balanced off by negative earnings if more than three disappointed offers occur). The mental health of patients, poor publicity arising from a large number of patients being disappointed (which could lead to trust issues with the institution, producing a future decrease in donation rates), and time-value of healthcare providers who participate in the decision process are very important concerns; and these are the reasons we considered such a high disappointment cost. Nonetheless, our model is flexible enough to easily accommodate changes in the cost parameters, which makes our model applicable for different economic conditions, and also realistic, as different values can be used for different organs. As part of a sensitivity analysis, if the disappointment cost is increased then the number of offers is reduced, and so is the number of organs ultimately donated, which makes the choice of parameters relevant, and something that must be aligned with the vision of the respective entities (SRTR, OPTN)\@. Finally, it is notable that our simulation-optimization framework can be used to test different policies and conditions on the arrivals of organs and patients.  Therefore, it may be worthwhile to undertake future work on extending the possible policies to test, or also to create a new ``gain'' function that better represents the trade-offs inherent in the transplantation system (which may be nonlinear).

\footnotesize
\section*{ACKNOWLEDGMENTS}
We  thank the OPTN and SRTR for facilitating private, high-quality data over the course of this project.  The data reported here have been supplied by the Hennepin Healthcare Research Institute (HHRI) as the contractor for the Scientific Registry of Transplant Recipients (SRTR)\@. The interpretation and reporting of these data are the responsibility of the author(s) and in no way should be seen as an official policy of or interpretation by the SRTR or the U.S. Government. We thank Alex Stroh for preliminary discussions.

\footnotesize

\bibliographystyle{wsc}

\bibliography{demobib2}

\begin{thebibliography}{}

\bibitem[\protect\citeauthoryear{Bertsimas, Farias, and Trichakis}{Bertsimas
  et~al.}{2013}]{[8]}
Bertsimas, D., V.~F. Farias, and N.~Trichakis. 2013.
\newblock ``Fairness, Efficiency, and Flexibility in Organ Allocation for
  Kidney Transplantation''.
\newblock {\em Operations Research\/}~61(1):73--87.


\bibitem[\protect\citeauthoryear{Blok, Detry, Putter, Rogiers, Porte, van Hoek,
  Pirenne, Metselaar, Lerut, Ysebaert, Lucidi, Troisi, Samuel, den Dulk,
  Ringers, Braat, and Committee}{Blok et~al.}{2016}]{[13]}
Blok, J.~J., O.~Detry, H.~Putter, X.~Rogiers, R.~J. Porte, B.~van Hoek,
  J.~Pirenne, H.~J. Metselaar, J.~P. Lerut, D.~K. Ysebaert, V.~Lucidi, R.~I.
  Troisi, U.~Samuel, A.~C. den Dulk, J.~Ringers, A.~E. Braat, and E.~L. I.~A.
  Committee. 2016.
\newblock ``Longterm Results of Liver Transplantation from Donation After
  Circulatory Death''.
\newblock {\em Liver Transplantation\/}~22(8):1107--1114.


\bibitem[\protect\citeauthoryear{Cabello, Rodriguez-Benot, Mazuecos, Osuna, and
  Alonso}{Cabello et~al.}{2011}]{[16]}
Cabello, M., A.~Rodriguez-Benot, A.~Mazuecos, A.~Osuna, and M.~Alonso. 2011.
\newblock ``Impact of Cold Ischemia Time on Initial Graft Function and Survival
  Rates in Renal Transplants from Deceased Donors Performed in Andalusia''.
\newblock {\em Transplantation Proceedings\/}~43(6):2174--2176.


\bibitem[\protect\citeauthoryear{Cao, Shahrestani, Chew, Crawford, Macdonald,
  Laurence, Hawthorne, Dhital, and Pleass}{Cao et~al.}{2016}]{[14]}
Cao, Y., S.~Shahrestani, H.~C. Chew, M.~Crawford, P.~S. Macdonald, J.~Laurence,
  W.~J. Hawthorne, K.~Dhital, and H.~Pleass. 2016.
\newblock ``Donation After Circulatory Death for Liver Transplantation: A
  Meta-Analysis on the Location of Life Support Withdrawal Affecting
  Outcomes''.
\newblock {\em Transplantation\/}~100(7):1513--1524.


\bibitem[\protect\citeauthoryear{{Health Resources and Services
  Administration}}{{Health Resources and Services Administration}}{2020}]{[2]}
{Health Resources and Services Administration} 2020.
\newblock ``{Organ Donation Statistics Summary, July 2020}''.
\newblock \url{https://www.organdonor.gov/statistics-stories/statistics.html},
  accessed 21.09.2021.

\bibitem[\protect\citeauthoryear{{International Registry of Organ Donation and
  Transplantation}}{{International Registry of Organ Donation and
  Transplantation}}{2019}]{[3]}
{International Registry of Organ Donation and Transplantation} 2019.
\newblock ``{IRODaT Database, 2019}''.
\newblock \url{https://www.irodat.org/?p=databasedata}, accessed 21.09.2021.

\bibitem[\protect\citeauthoryear{Kilinc, Shahraki, Degnim, Hoskin, Horton, Sir,
  Pasupathy, and Gel}{Kilinc et~al.}{2020}]{Kilinc2020}
Kilinc, D., N.~Shahraki, A.~C. Degnim, T.~L. Hoskin, T.~M. Horton, M.~S. Sir,
  K.~S. Pasupathy, and E.~S. Gel. 2020.
\newblock ``Simulation Modeling as a Decision Tool for Capacity Allocation in
  Breast Surgery''.
\newblock In {\em Proceedings of the 2020 Winter Simulation Conference}, edited
  by\ K.-H.~G. Bae, B.~Feng, S.~Kim, S.~Lazarova-Molnar, Z.~Zheng, T.~Roeder,
  and R.~Thiesing,  806–817.
\newblock Piscataway, New Jersey: Institute of Electrical and Electronics
  Engineers, Inc.

\bibitem[\protect\citeauthoryear{Konrad}{Konrad}{2020}]{Konrad2020}
Konrad, K. 2020.
\newblock ``A Simulation Model for the Multi-Period Kidney Exchange
  Incentivization Problem''.
\newblock In {\em Proceedings of the 2020 Winter Simulation Conference}, edited
  by\ K.-H.~G. Bae, B.~Feng, S.~Kim, S.~Lazarova-Molnar, Z.~Zheng, T.~Roeder,
  and R.~Thiesing,  794–805.
\newblock Piscataway, New Jersey: Institute of Electrical and Electronics
  Engineers, Inc.

\bibitem[\protect\citeauthoryear{Mankowski, Kosztowski, Raghavan,
  Garonzik-Wang, Axelrod, Segev, and Gentry}{Mankowski et~al.}{2019}]{[10]}
Mankowski, M.~A., M.~Kosztowski, S.~Raghavan, J.~M. Garonzik-Wang, D.~Axelrod,
  D.~L. Segev, and S.~E. Gentry. 2019.
\newblock ``Accelerating Kidney Allocation: Simultaneously Expiring Offers''.
\newblock {\em American Journal of Transplantation\/}~19(11):3071--3078.


\bibitem[\protect\citeauthoryear{Matesanz, Dominguez-Gil, Coll, Mahíllo, and
  Marazuela}{Matesanz et~al.}{2017}]{[15]}
Matesanz, R., B.~Dominguez-Gil, E.~Coll, B.~Mahíllo, and R.~Marazuela. 2017.
\newblock ``How Spain Reached 40 Deceased Organ Donors per Million
  Population''.
\newblock {\em American Journal of Transplantation\/}~17(6):1447--1454.


\bibitem[\protect\citeauthoryear{Moustaid, Kornevs, and Meijer}{Moustaid
  et~al.}{2019}]{Moustaid2019}
Moustaid, E., M.~Kornevs, and S.~Meijer. 2019.
\newblock ``Sensitivity Analysis of Policy Options for Urban Mental Health
  Using System Dynamics and Fuzzy Cognitive Maps''.
\newblock In {\em Proceedings of the 2019 Winter Simulation Conference}, edited
  by\ N.~Mustafee, K.-H.~G. Bae, S.~Lazarova-Molnar, M.~Rabe, C.~Szabo,
  P.~Haas, and Y.-J. Son,  1055–1066.
\newblock Piscataway, New Jersey: Institute of Electrical and Electronics
  Engineers, Inc.

\bibitem[\protect\citeauthoryear{{Organ Procurement and Transplantation
  Network}}{{Organ Procurement and Transplantation Network}}{2020}]{[1]}
{Organ Procurement and Transplantation Network} 2020.
\newblock ``{Data Summary, July 2020}''.
\newblock \url{https://optn.transplant.hrsa.gov/data/}, accessed 21.09.2021.

\bibitem[\protect\citeauthoryear{Pan, Yoeli, Galvan, Kueht, Cotton, O'Mahony,
  Goss, and Rana}{Pan et~al.}{2018}]{[11]}
Pan, E.~T., D.~Yoeli, N.~Galvan, M.~L. Kueht, R.~T. Cotton, C.~A. O'Mahony,
  J.~A. Goss, and A.~Rana. 2018.
\newblock ``Cold Ischemia Time is an Important Risk Factor for Post-Liver
  Transplant Prolonged Length of Stay''.
\newblock {\em Liver Transplantation\/}~24(6):762--768.


\bibitem[\protect\citeauthoryear{Peters-Sengers, Houtzager, Idu, Heemskerk, van
  Heurn, Homan van~der Heide, Kers, Berger, van Gulik, and
  Bemelman}{Peters-Sengers et~al.}{2019}]{[12]}
Peters-Sengers, H., J.~Houtzager, M.~M. Idu, M.~Heemskerk, E.~van Heurn, J.~J.
  Homan van~der Heide, J.~Kers, S.~P. Berger, T.~M. van Gulik, and F.~J.
  Bemelman. 2019.
\newblock ``Impact of Cold Ischemia Time on Outcomes of Deceased Donor Kidney
  Transplantation: An Analysis of a National Registry''.
\newblock {\em Transplantation Direct\/}~5(5):e448.


\bibitem[\protect\citeauthoryear{Sandik\c{c}i, Tun\c{c}, and
  Tanri\"{o}ver}{Sandik\c{c}i et~al.}{2019}]{Sandi2019}
Sandik\c{c}i, B., S.~Tun\c{c}, and B.~Tanri\"{o}ver. 2019.
\newblock ``A New Simulation Model for Kidney Transplantation in the United
  States''.
\newblock In {\em Proceedings of the 2019 Winter Simulation Conference}, edited
  by\ N.~Mustafee, K.-H.~G. Bae, S.~Lazarova-Molnar, M.~Rabe, C.~Szabo,
  P.~Haas, and Y.-J. Son,  1079–1090.
\newblock Piscataway, New Jersey: Institute of Electrical and Electronics
  Engineers, Inc.

\bibitem[\protect\citeauthoryear{{Scientific Registry of Transplant
  Recipients/Organ Procurement and Transplantation Network}}{{Scientific
  Registry of Transplant Recipients/Organ Procurement and Transplantation
  Network}}{2019}]{[4]}
{Scientific Registry of Transplant Recipients/Organ Procurement and
  Transplantation Network} 2019.
\newblock ``{Data Reports, 2019}''.
\newblock
  \url{https://www.srtr.org/reports-tools/srtroptn-annual-data-report/},
  accessed 21.09.2021.

\bibitem[\protect\citeauthoryear{Stahl, Kreke, Malek, Schaefer, and
  Vacanti}{Stahl et~al.}{2008}]{[17]}
Stahl, J.~E., J.~E. Kreke, F.~A. Malek, A.~J. Schaefer, and J.~Vacanti. 2008.
\newblock ``Consequences of Cold-Ischemia Time on Primary Nonfunction and
  Patient and Graft Survival in Liver Transplantation: a Meta-Analysis''.
\newblock {\em PLoS One\/}~3(6):e2468.


\bibitem[\protect\citeauthoryear{{United Network for Organ Sharing}}{{United
  Network for Organ Sharing}}{2018}]{[6]}
{United Network for Organ Sharing} 2018.
\newblock ``{Improving Organ Placement Processes in DonorNet}''.
\newblock
  \url{https://unos.org/news/improving-organ-placement-processes-in-donornet/},
  accessed 21.09.2021.

\bibitem[\protect\citeauthoryear{{United Network for Organ Sharing}}{{United
  Network for Organ Sharing}}{2019}]{[5]}
{United Network for Organ Sharing} 2019.
\newblock ``{10 Things UNOS is Doing to Increase Organ Utilization}''.
\newblock
  \url{https://unos.org/news/improvement/10-things-unos-is-doing-to-increase-organ-utilization/},
  accessed 21.09.2021.

\bibitem[\protect\citeauthoryear{Wolfe, LaPorte, Rodgers, Roys, Fant, and
  Leichtman}{Wolfe et~al.}{2007}]{[9]}
Wolfe, R.~A., F.~B. LaPorte, A.~M. Rodgers, E.~C. Roys, G.~Fant, and A.~B.
  Leichtman. 2007.
\newblock ``Developing Organ Offer and Acceptance Measures: When `Good' Organs
  are Turned Down''.
\newblock {\em American Journal of Transplantation\/}~7:1404--1411.


\bibitem[\protect\citeauthoryear{Zenios, Chertow, and Wein}{Zenios
  et~al.}{2000}]{[7]}
Zenios, S.~A., G.~M. Chertow, and L.~M. Wein. 2000.
\newblock ``Dynamic Allocation of Kidneys to Candidates on the Transplant
  Waiting List''.
\newblock {\em Operations Research\/}~48(4):549--569.


\end{thebibliography}

\section*{AUTHOR BIOGRAPHIES}

\noindent {\bf IGNACIO ERAZO} is a third-year Ph.D. student majoring in Operations Research and minoring in Machine Learning in the H. Milton Stewart School of Industrial and Systems Engineering at the Georgia Institute of Technology. His research interests include the development of efficient optimization algorithms and heuristics for intelligent decision-making, as well as large-scale simulation-optimization procedures. His e-mail address is \email{ierazo@gatech.edu}, and his website is \href{https://sites.google.com/view/ignacio-erazo}{https://sites.google.com/view/ignacio-erazo}.\\[-3pt]

\noindent {\bf DAVID GOLDSMAN} is a Professor in the H. Milton Stewart School of Industrial and Systems Engineering at the Georgia Institute of Technology. Dave's research interests include simulation output analysis, statistical ranking and selection methods, and humanitarian applications. His email address is \email{sman@gatech.edu}, and his website is \href{https://www.isye.gatech.edu/~sman/}{https://www.isye.gatech.edu/$\sim$sman/}. \\[-3pt]

\noindent {\bf PINAR KESKINOCAK} is the William W. George Chair and Professor in the H. Milton Stewart School of Industrial and Systems Engineering at Georgia Tech. Dr.\ Keskinocak's research focuses on the applications of operations research and management science with societal impact, particularly health and humanitarian applications, supply chain management, and logistics/transportation. Her email address is \email{pinar@isye.gatech.edu}, and her website is \href{https://www2.isye.gatech.edu/people/faculty/Pinar_Keskinocak/}{https://www2.isye.gatech.edu/people/faculty/Pinar\_Keskinocak/}. \\[-3pt] 

\noindent {\bf JOEL SOKOL} is a Professor in the H. Milton Stewart School of Industrial and Systems Engineering at Georgia Tech. His research focuses on healthcare systems, analytics, and applied operations research. His email address is \email{jsokol@isye.gatech.edu}.

\end{document}